\documentstyle[11pt,newpasp,twoside,epsf]{article}
\def\edcomment#1{\iffalse\marginpar{\raggedright\sl#1\/}\else\relax\fi}
\reversemarginpar

\begin{document}

\title{Radio Properties of REX BL Lacs and Galaxies}
\author{Anna Wolter$^{1}$, A. Caccianiga$^{2}$, T. Maccacaro$^{1}$, 
R. Della Ceca$^{1}$, I.M. Gioia$^{3,4}$, F. Cavallotti$^{5}$, 
M. Minoia$^{5}$ }
\affil{$^{1}$Osservatorio Astronomico di Brera, Milano, Italy\\
$^{2}$Observat\'orio Astron\'omico de Lisboa, Lisboa, Portugal \\ 
$^{3}$Istituto di Radioastronomia, Bologna, Italy \\
$^{4}$Institute for Astronomy, Honolulu, HI \\
$^{5}$Universit\`a degli Studi di Milano, Milano, Italy}

\begin{abstract} 
Detailed VLA observations have been gathered for a number of sources 
classified as either BL Lacs or galaxies, derived from the REX survey.
We focus in particular on the sources identified by us, for which we have 
in hand homogeneous optical data, to study in more detail than allowed by 
the NVSS the radio properties of these 
sources in the framework of AGN unified models. 
\end{abstract}

\section{Introduction}
The REX survey (Radio Emitting X-ray sources; see Maccacaro et al. 1998,
Caccianiga et al. 1999 for details of the survey) is based on pointed 
ROSAT/PSPC observations and NVSS (NRAO VLA Sky Survey, Condon et al. 1998) 
data. 
About 500 of the $\sim 1600$ REX sources have an optical identification 
obtained working down from the highest 
X-ray fluxes, in order to have larger and larger completely identified
subsamples  (see Caccianiga et al., these proceedings for the first results
on the Xray--Bright-REX sample of BL Lacs).

\begin{table}
\caption{Identification breakdown as of July 2000}
\begin{center}
\begin{tabular}[h]{|l r r r |}
\hline
ID 	 & New  & Literature & Total \\
AGN 	 & 96  	& 138  & 234	 \\
BL+cand. & 40 	& 32   & 72	 \\
Galaxies & 55 	& 121  & 176	 \\
\hline
\end{tabular}
\end{center}
\label{id}
\end{table}

AGN and BL Lacs are the most common counterparts, followed by galaxies, 
either isolated or in clusters.
The objects without emission lines are classified according to the 
amount of the CaII discontinuity in our optical 
spectral data.
Values larger than 40\% define galaxies. BL Lacs are defined as 
having a CaII discontinuity smaller than 25\%,
while objects with CaII discontinuity between 25\% and 40\% are defined
as BL Lac candidates.
Most of the galaxy classification from the literature are instead based on
the optical extended morphology.
In Table 1 
we list the updated identification breakdown.

\section{Testing unified models}

Unified models (see Urry \& Padovani 1995 for a review) imply 
that both radio and optical appearance
are governed by the proximity of the line of sight to the radio axis, 
which is supposed to correspond also to the rotation axis of the accretion disk.

We want to test if the angle to the line of sight is really the primary
parameter on which the transition between galaxies and BL Lacs rests.
To do so we concentrate on the subsample of non-emission line objects
(i.e galaxies, BL Lacs and candidates), in particular those that have been
observed by us, since for those objects we have homogeneous optical data.
In the radio band, we expect a transition from an extended steep-spectrum
source to a dominant flat-spectrum core due to the Doppler boosted foreground
jet. In the optical band the direction with respect to the line of
sight should be measured by a  CaII  absorption less and less prominent
as the angle decreases, due to the increasing dominance of the non-thermal
(synchrotron) component.

We have thus performed VLA observations at two frequencies (20 and 6 cm) in two
different configurations. Forty-one objects have been observed in the most
resolved configuration (A)
and a subset of 35 in B configuration.
The objects are almost equally divided in Galaxies (11), BL Lac candidates
(12) and ``bona-fide" BL Lacs (18), according, as said before, to the
measure of the CaII depression.
The resolution of these observations ranges from $\sim0.3^{\prime\prime}$ to
$\sim3^{\prime\prime}$, depending on the VLA configuration and observing
frequency.
At the redshift of the observed sample (z $\sim 0.1-0.2$) we are therefore
probing structures of the order of a few kpc.

\subsection{Core dominance}

\begin{figure}
\plottwo{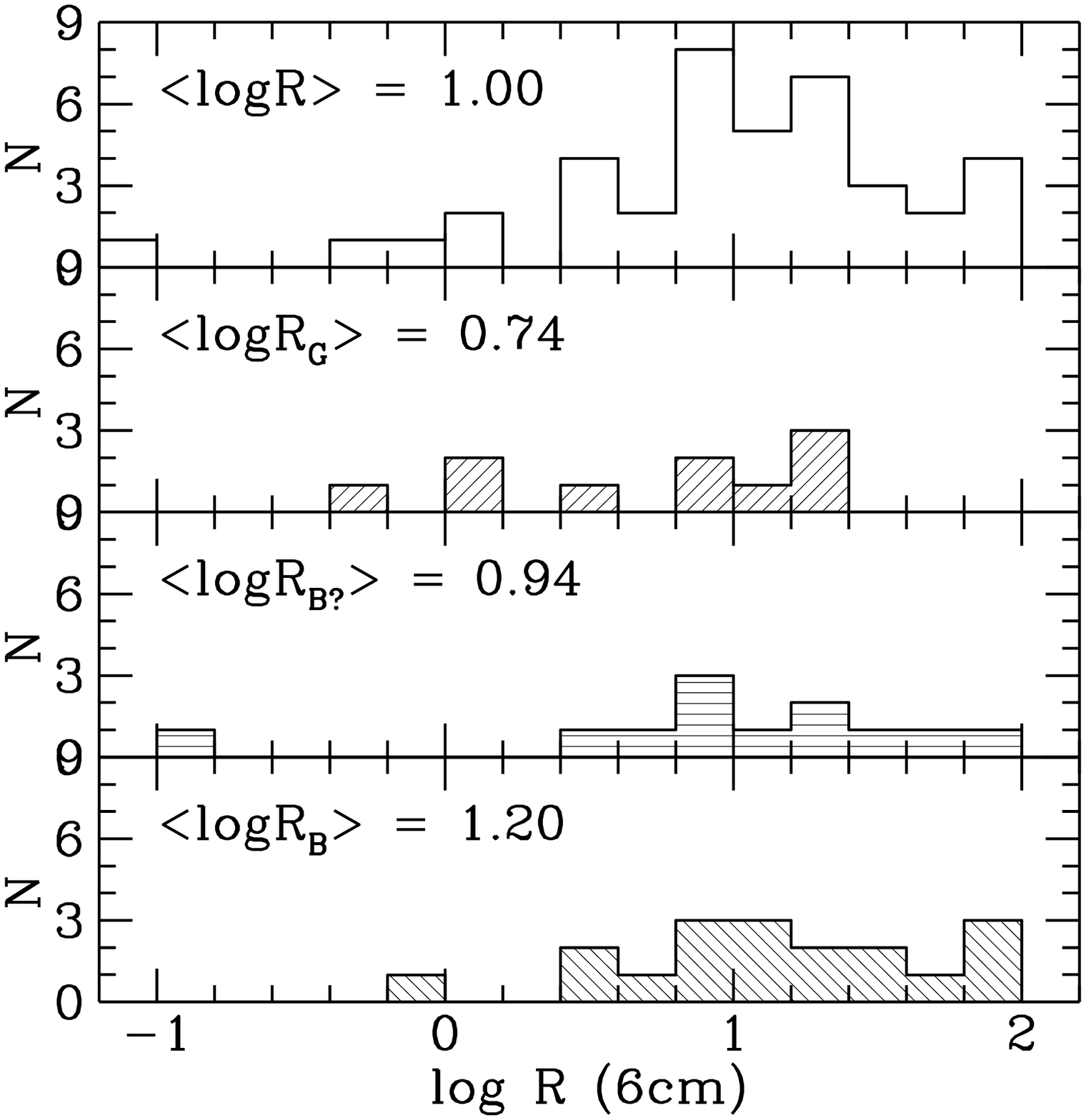}{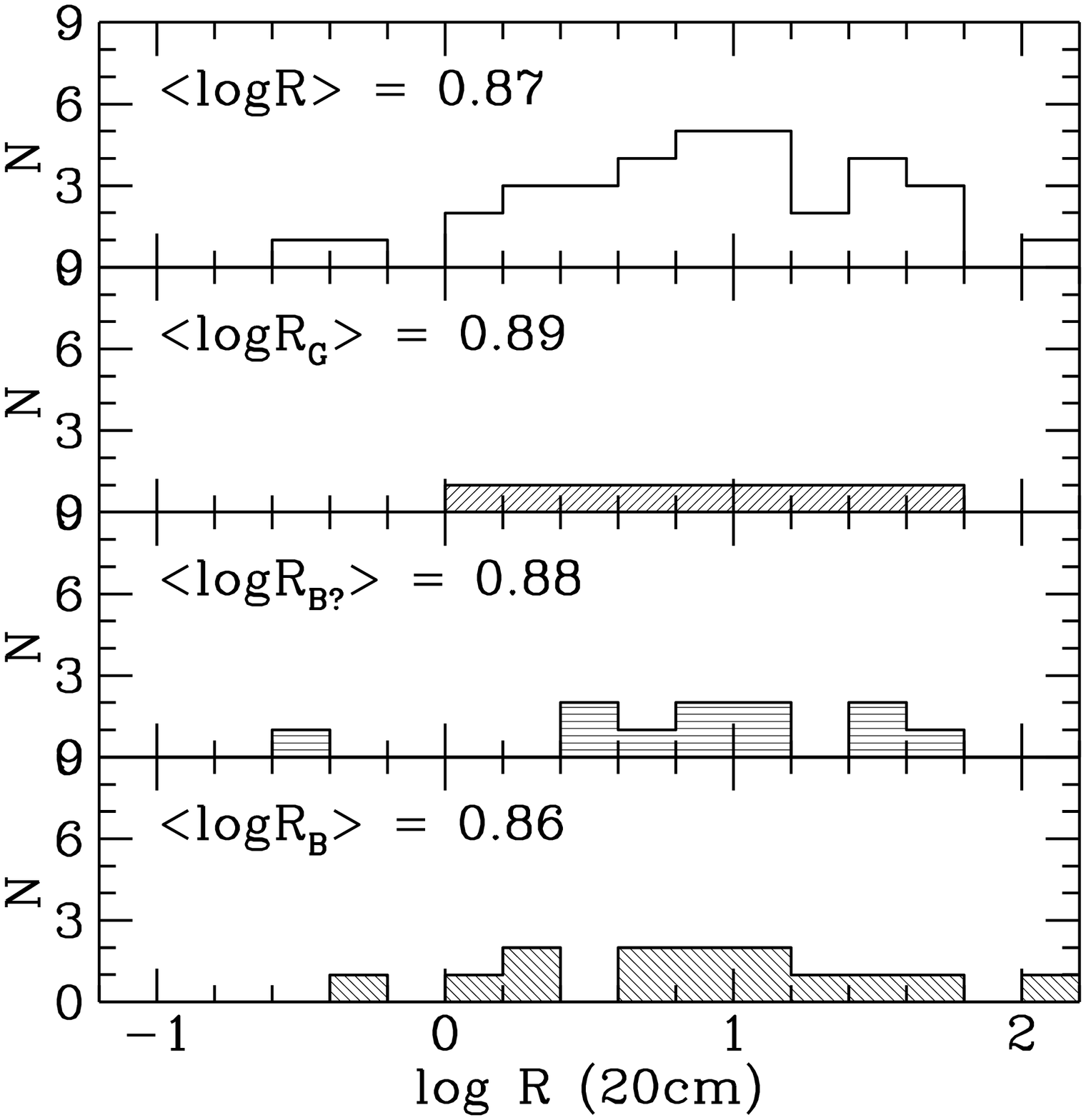}
\caption{{\it Left:} The core dominance measured at 6 cm in configuration A
-- i.e. from the most resolved data. 
From top to bottom: the total sample, the galaxies, the BL Lac candidates
and the ``bona-fide" BL Lacs. The mean values for each subsamples are 
listed in each panel. {\it Right:} The core dominance measured at 20 cm -- 
as is commonly found in literature. 
}
\end{figure}

One of the best indicators of the orientation of the beamed radiation
with respect to the line of sight in the radio band is the so-called
``core dominance", $R$, defined as the ratio between the {\it core} flux
($F_C$) and the {\it extended} flux. In practice, the core flux is defined as
the peak of the emission, while the extended flux is the difference between
the total integrated flux ($F_T$) and the peak flux:
 $R = { F_C \over {(F_T - F_C)} }$

We compute the core dominance $R$ at both 6cm and 20cm for the A configuration
data. The first frequency
gives the best resolution available in our data, however in many literature 
papers -- from the VLA and comparable size telescopes --  the 20cm data are
used.

In Figure 1 
we plot the distribution of the core dominance 
at the two frequencies for the different subsamples
and for the total sample. Mean values are also reported in the figure.
We can compare them with values found in the literature (Morganti et al. 1997;
Laurent-Muehleisen et al. 1993) for analogous samples
of objects. We have: 
$\langle$logR$\rangle$=--1.6 (for a sample of classical FRI); 
$\langle$logR$\rangle$=0.04 (for a sample of FRI galaxies, with an 
extension to fainter objects from the B2 survey); 
$\langle$logR$\rangle$=0.05 (Steep Spectrum Quasars); 
$\langle$logR$\rangle$=0.5 (1Jy BL Lacs);
$\langle$logR$\rangle$=1.28 (Flat Spectrum Quasars); 
$\langle$logR$\rangle$=1.3 (EMSS and HEAO 1 LASS BL Lacs).
The fact that different values of $R$ are found for the same objects
by using the data sets at the two frequencies implies that caution 
should be taken when comparing different resolution data.

\begin{figure}
\plotfiddle{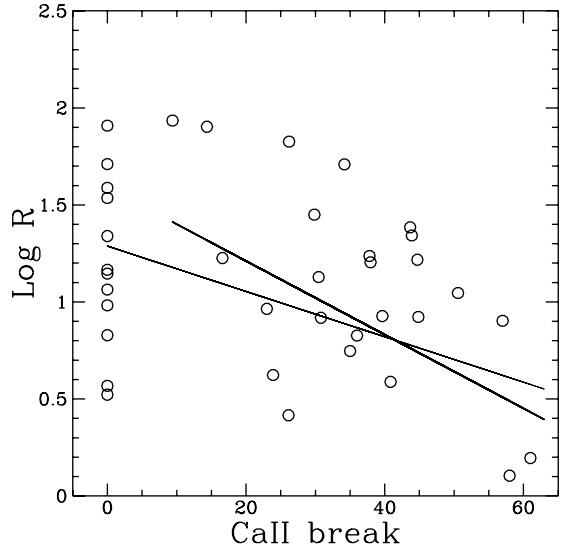}{8cm}{0}{38}{38}{-140}{-40}
\vskip -1cm
\caption{The correlation between the core dominance from VLA data
at 6cm and the CaII depression as measured from optical data. The two
lines are regression fits, see text for details}
\label{corr}
\end{figure}

We can explore a larger sample by using catalog data, in particular
the FIRST and NVSS data, although not simultaneous, to compute the core
dominance for a subset of REX galaxies, BL Lacs and candidates that have
not been observed by us at the VLA.
We derive somewhat lower values of $R$, since we include in this set also
some of the well-known nearby objects.
In any case, in this larger subsample there is a trend between isolated 
galaxies, galaxies in clusters,
and BL Lacs ($\langle$logR$\rangle$ = --0.04; 0.28; 0.86, respectively), 
possibly due to
an environment effect because the denser environment in a cluster could 
constrain the radio emitting gas.
However, the overlap in the distribution is large, and the radio galaxies 
in the REX are definitely more core-dominated than classical FRI.

If $R$ and the CaII depression are indicators
of orientation, then we expect to find a strong correlation
between the two. We plot such function in Figure 2. 
As evident from the figure, the trend in $R$ with CaII break
is very marginal
(regression coefficient = --0.4) and also the slope of the regression fit
is very flat: we obtain --0.01 (thin line) or --0.02 (thick line) 
by including or excluding the objects 
in which the CaII depression is not visible in the spectrum, that are 
set at 0.

\subsection {Radio Morphology}

\begin{figure}
\plotfiddle{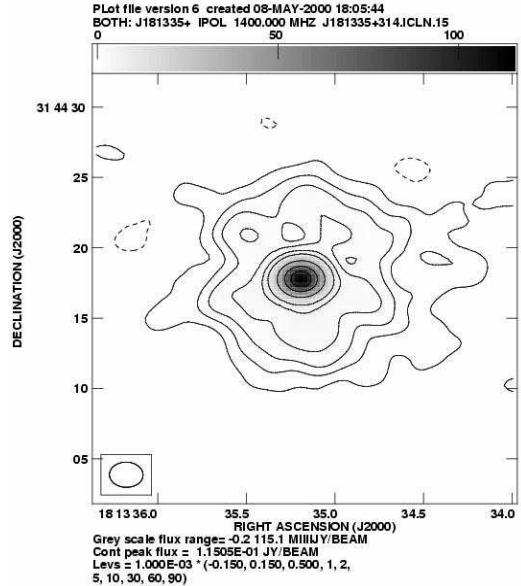}{10cm}{0}{85}{85}{-120}{50}
\vskip -2cm
\caption{The 20cm radio map of 1REXJ181335+3144. The halo around the
central pointlike source is only a few percent of the total flux.}
\end{figure}

We expect a more complex radio morphology  for objects that preferentially
lie in the plane of the sky. It was therefore a bit of a surprise to find that
most of the resolved objects are classified as BL Lacs or candidates, and
not as galaxies. Actually only $\sim$10\% of the objects have been detected with
an extended structure in the most resolved configuration, and so the number
of objects under consideration is
very small. Furthermore, one of the resolved object, 1REXJ181335+3144,
is consistent with
the picture of being the core of a BL Lac jet pointed at us, surrounded by
a very faint halo, due to the lobe seen in projection (see Figure 3).

\begin{figure}
\plottwo{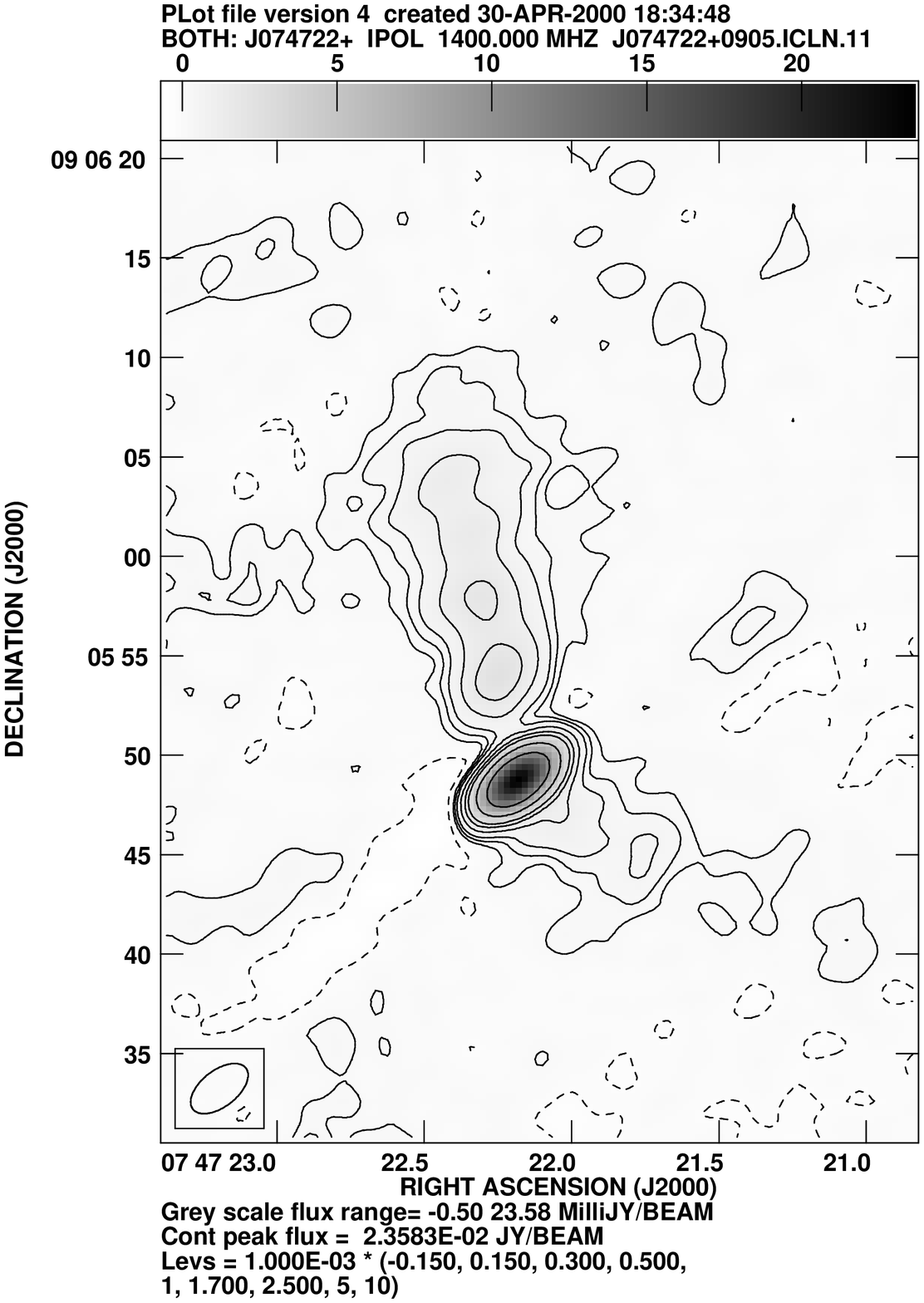}{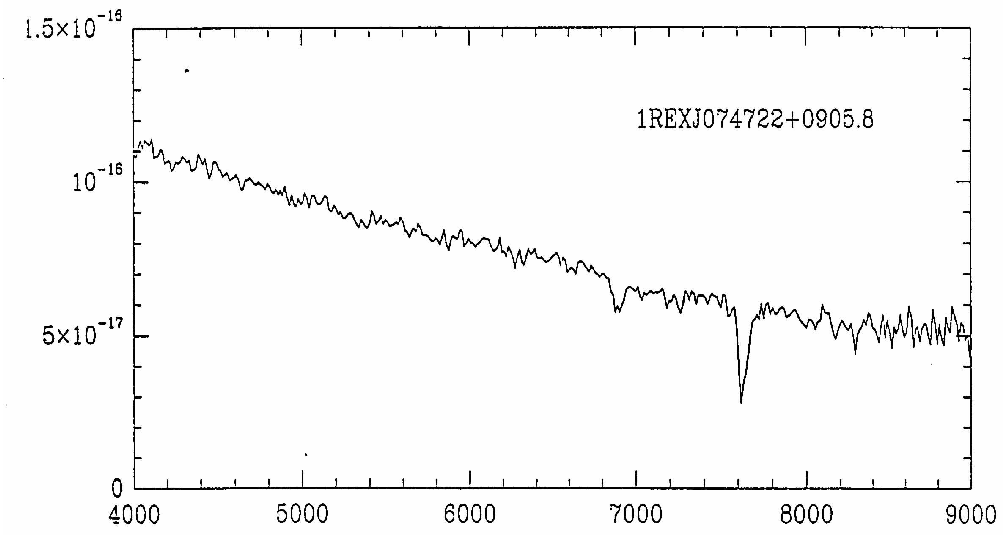}
\caption{{\it Left:} The 20cm radio map of 1REXJ074722+0905; {\it Right:}
the featureless optical continuum of the same source }
\end{figure}

The most striking and unexpected morphology comes from the object shown
in Figure 4. 
This object, 1REXJ074722+0905, as can be seen from the featureless optical 
continuum, is classified as a BL Lac, with no redshift determination.
The prominent structure observed is $\sim 20^{\prime\prime}$ long.
The radio spectral index (defined by $F_{\nu} \sim \nu^{\alpha}$) 
is --0.6 for the total source,
and --0.3 for the core, implying that the extended part is indeed steeper.
The extent was already suggested by the NVSS data (with a restoring beam of 
$45^{\prime\prime}$),  where the source, although
described by a single Gaussian component, has an integrated flux about
10\% larger than the peak flux. 
Similar morphologies have been found for a few 1Jy BL Lacs (Cassaro et al. 
1999), in particular for 1147+245, for which no emission or absorption lines 
are seen in the spectrum as in 1REXJ074722+0905. 
The other 1Jy objects with extended radio structure
show both absorption and narrow emission lines in the optical spectrum.

\section{Conclusions}

From the REX survey we have extracted a number of BL Lacs and galaxies,
to study their radio properties in the context of Unified Models.
Our results seem to imply that objects classified spectroscopically as
galaxies, i.e. that do not have emission lines and have a large predominance
of stellar light, as measured form the CaII absorption, are not the same 
objects that are defined as classical radio galaxies, i.e. with a large
radio structure of jets and lobes in the plane of the sky.
The resolved morphology contrasts with the BL Lac classification from the
optical spectrum in at least one striking case.

The two measures of orientation: core domninance and CaII depression, 
from the radio and optical band respectively,
seem to be only weakly correlated, suggesting that a revision of the
unifying models, in which not only the orientation contributes to change
one class into the other, might be necessary.

\begin{acknowledgements}
It is a pleasure to thank Greg Taylor for his precious help in 
reducing the VLA data. 
This work has received partial financial support from 
the Italian Space Agency (ASI) and the Italian MURST under grant
COFIN98-02-32. 
\end{acknowledgements}


\begin{references}

\reference Caccianiga, A., et al. 1999, ApJ, 513, 51

\reference Cassaro, P., Stanghellini, C., Bondi, M., Dallacasa, D.,
 Della Ceca, R., and Zappal\`a, R.A., 1999, A\&AS, 139, 601.

\reference Condon, J.J., et al., 1998, AJ, 115, 1693.
 
\reference Laurent-Muehleisen, S.A., Kollgaard, R.I., Moellenbrock, G.A.,
 Feigelson, E.D., 1993, AJ, 106, 875.

\reference Maccacaro, T., et al. 1998, Astron. Nachr. 319, 15

\reference Morganti, R., Oosterloo, T.A., Reynolds, J.E., Tadhunter, C.N.,
 Migenes, V., 1997, MNRAS, 284, 541.

\reference Urry, C.M. \& Padovani, P., 1995, PASP, 107, 803.
\end{references}
\end{document}